# Eight Minutes and a Half

**Gabriele U. Varieschi,** Loyola Marymount University, Los Angeles, CA

"It takes about eight minutes and a half for the light from the Sun to reach us, therefore when we look at a beautiful sunset and we see the Sun at the horizon… the Sun is actually not there anymore, it's already below the horizon! The real sunset happened eight and a half minutes earlier! Similarly, at sunrise, the Sun seems to be at the horizon, but is already up in the sky, due to the same time delay."

I first heard this statement from a friend, a former colleague in physics teaching who found it in different publications, ranging from physics textbooks, general science books and astronomical magazines.[1]

At first, the quoted statement seems perfectly reasonable. After all if something happens on the Sun, for example if new sunspots were to develop on its surface or if for some strange reason the Sun suddenly were to turn green or purple, we would actually observe these events with a time delay of approximately eight and a half minutes.

Or one can think of apparently similar situations, such as observing the lights of a moving car at night, with the car suddenly turning around a corner and disappearing from sight. The time delay for the light to reach us would be practically minimal, but the car is effectively already around the corner when its last light reaches us.

A quick informal poll of some of my students revealed that most of them tend to agree with the statement, for reasons similar to those I just mentioned. However, a closer inspection of the statement reveals the real hidden issue: a problem of rotating vs. inertial reference systems. In this paper I will analyze this problem and describe a simple experimental activity which can also be used as a demonstration of the peculiar kinematical effects of rotating systems.

**The description of the problem**

To simplify the problem, we can avoid unnecessary complications by taking the Sun to be a point-like source of radiation, so that there is no ambiguity in the meaning of *sunrise*[2] or *sunset*: these events happen when we observe the point-like Sun exactly at the horizon, i.e., along a geometrical tangent line to the Earth's surface, at our point of observation.

Let's remove also any optical effect from the problem: no light refraction in the atmosphere (let's remove the atmosphere altogether, if we prefer), no other light bending of any sort due to gravitational fields. We can simply assume that the radiation from the Sun travels in perfectly straight lines at the speed of light in vacuum.

Sunset and sunrise are essentially due to the daily rotation of the Earth around its axis, with a period we can take to be exactly 24 hours. Let's assume that the time delay for the light to reach us from the Sun is exactly 8 minutes (we don't need the extra few seconds).

An elementary calculation shows that in 8 minutes the Earth rotates an angle of 2 degrees, equivalent to an *apparent* rotation of the Sun by the same amount in the opposite direction. If the statement quoted at the beginning is correct, we would conclude that at sunset the Sun is already below the horizon by an angle of two degrees, a quite significant change in position.

The orbital revolution of the Earth around the Sun will also affect the angle of rotation, but its effect is minimal compared to the one just discussed and will be neglected here. After all, if the Earth did not rotate around its axis, our "day" would last one full year and the apparent rotation of the Sun in eight minutes would amount to less than 20 seconds of a degree, a very minimal correction compared to the 2 degrees calculated above.

Therefore, we can simply assume a fixed Sun, behaving like a point-like source of radiation, with the emitted light traveling in straight lines and consider the Earth as a rotating sphere, with a period of 24 hours, receiving light from the Sun, with the photons taking about 8 minutes to complete their trip.

At this point every physics student should know how to interpret the problem. The "fixed" Sun reference frame can be assumed to be an inertial one; the Sun is not moving and is emitting "projectiles" (photons) which travel radially outward at constant speed. The Earth represents a rotating, non-inertial system; therefore we should be careful when observing and interpreting phenomena in such non-inertial systems.

The situation as seen from the inertial system of the Sun is extremely simple and is illustrated in Figs. 1a and 1b. In this reference system (x-y-z axis), the position of the Sun (S) is fixed, while the Earth observer (point T) is rotating together with the local direction of the horizon, indicated in the figures by the blue dashed line, tangent to the Earth's surface at point T. We choose to place the terrestrial observer T on the equator for simplicity, but any other latitude would give the same results, just with a radius R smaller than the equatorial radius (of course at latitudes beyond the Arctic or Antarctic Circles we might have to wait a little longer than usual to observe a new sunrise or sunset). These figures show the view from the North Pole, i.e., with the Earth rotating counter-clockwise around the z axis (perpendicular to the plane of the figures). We also indicate with x'-y'-z' the rotating system, where the z and z' axis (coming out of the page) coincide with the axis of rotation.

Fig. 1a shows the situation eight minutes before "sunset": at this time the last rays to be observed on Earth depart from the Sun, but the Sun is still well above the horizon. In fact, the angle between the horizon direction at point T and the direction of arrival of sunlight (dashed red line) corresponds to the 2 degrees angle calculated above (and greatly exaggerated in Fig. 1a).

In Fig. 1b we illustrate the "sunset" situation. The light rays emitted eight minutes earlier finally arrive to the observer at position T and their direction of arrival is aligned perfectly with the horizon direction: at sunset the Sun position is precisely "at the horizon" and not below it! Similar reasoning would obviously apply for sunrise events, but we prefer to continue doing all our examples just with sunsets.

Although the original statement already appears to be incorrect, one might argue that it was referring to an observer on Earth, so we should analyze the problem in the rotating reference system instead and illustrate the situation also from this point of view.

Fig. 2 shows the path of the "last" light from the Sun as seen from the observer rotating with the Earth. From this perspective the apparent rotational motion of the Sun (with a 24 hour period) is a clockwise motion, when viewed from the North Pole.

For the non-inertial observer at position T, the path followed by the light is not straight anymore, but is a spiral-like curve starting from the original position of the Sun (point A), eight minutes before we observe sunset, and ending with light reaching point T eight minutes later, with a direction of motion (given by the tangent to the path at point T) coinciding with the "horizon" direction, so that in our rotating system we perceive the light at sunset as coming straight towards us from the horizon.

And where is the Sun at sunset, from our rotating perspective? In those eight minutes it took the light to reach us at sunset, the Sun "moved" clockwise from position A to position B, as seen in Fig. 2, so it is exactly at the horizon when we receive the last light. Any light emitted from the Sun at position B (or from any intermediate position between A and B) would not reach the observer at position T any more. It would have to follow a similar curved path reaching the Earth with a final direction of motion "below" the horizon, therefore would not be seen by the observer at T.

In any case, from both perspectives, the Sun is exactly at the horizon at the precise moment of sunset or sunrise.

**A more analytical approach**

The analysis of the previous section can be made more quantitative by using rotating coordinate systems. As illustrated in Fig. 1a and 1b, we consider the Sun as point-like source (S) located at a distance D of one astronomical unit, $D = 1\,AU = 1.496 \times 10^{11}\,m$, and our planet as a sphere of radius $R = 6.378 \times 10^6\,m$, equal to the Earth's equatorial radius. The ratio $R/D = 4.26 \times 10^{-5}$ is therefore very small and our figures are obviously not up to scale.

For our problem it is convenient to use a very particular set of units: distances will be measured in astronomical units (AU), time in minutes (min) and angles in degrees (°). Assuming that the time needed for light to travel the Sun-Earth distance is exactly eight minutes, the speed of light c and the angular velocity ω of our planet can be expressed by very simple numbers:[3]

$$c = \tfrac{1}{8}\tfrac{AU}{\min};\; \omega = \tfrac{1}{4}\tfrac{°}{\min}, \tag{1}$$

so that, as already remarked, in eight minutes the Earth rotates an angle of two degrees.

We will also take $t = 0$ to be the time of "sunset", i.e., when the Earth observer receives the last light at position T. This light left the Sun eight minutes earlier, therefore at the initial time $t_0 = -8\,\min$. At this time we will assume that the x'-y'-z' system is rotated by an initial angle $\alpha_0 = -2°$, with respect to the fixed x-y-z system, around the common z, z' axis. In this way, eight minutes later at time $t = 0$, the two systems will coincide at sunset. The uniform rotation of the primed system is thus described by the angle of rotation α:

$$\alpha = \alpha_0 + \omega(t - t_0) = -2 + \tfrac{1}{4}(t+8) \tag{2}$$

and the equation of motion of the light "projectiles" in the direction tangent to the Earth's surface, in the inertial system, describes a simple uniform motion in a straight line:

$$\begin{aligned} x &= D - c(t - t_0) = 1 - \tfrac{1}{8}(t+8) \\ y &= R \end{aligned} \tag{3}$$

when expressed in the particular units we have introduced above. With this choice of units we can compare the expressions in Eqs. (2) and (3) and find a useful relation between x and α:

$$x = -\tfrac{1}{2}\alpha. \tag{4}$$

At this point we can determine the equations of motion of light in the rotating system x'-y'-z' and show that it is indeed following the curved trajectory shown in Fig. 2. The connection between the primed and non primed coordinates is given by a simple rotation around the z, z' axis, by the angle α:

$$\begin{aligned} x' &= \cos\alpha\, x + \sin\alpha\, y = -\tfrac{1}{2}\alpha\cos\alpha + R\sin\alpha \\ y' &= -\sin\alpha\, x + \cos\alpha\, y = \tfrac{1}{2}\alpha\sin\alpha + R\cos\alpha \end{aligned} \tag{5}$$

where we used Eqs. (3) and (4) to express the coordinates x', y' as a function of a common parameter α. These parametric equations of motion, when plotted for α varying between $\alpha = -2°$ (the initial angle) and $\alpha = 0°$, will reproduce the curve in Fig. 2.

To plot this solution it is actually easier to consider plane polar coordinates r', φ':

$$\begin{aligned} r' &= \sqrt{x'^2 + y'^2} = \sqrt{\tfrac{1}{4}\alpha^2 + R^2} \\ \varphi' &= \arctan\frac{y'}{x'} = \arctan\left(\frac{\tfrac{1}{2}\alpha\sin\alpha + R\cos\alpha}{-\tfrac{1}{2}\alpha\cos\alpha + R\sin\alpha}\right) \end{aligned} \tag{6}$$

where again the parameter α varies from $-2°$ to $0°$. It is easy to check from Eq. (6) that at sunset ($\alpha = 0°$) the light beam reaches the Earth observer at T ($r' = R$ and $\varphi' = 90°$ for $\alpha = 0°$) and that its direction of arrival is aligned with the horizon direction ($\frac{dr'}{dt} = 0$ for $t = 0$).

Alternatively, one can use the Cartesian components of the velocity in the primed coordinates, obtained from the time derivatives of Eq. (5), to study the direction of motion at any point along the trajectory. It is also easy to check the direct connection between the velocities in the two systems, given by the classic relation $\vec{v} = \vec{v}' + \vec{\omega} \times \vec{r}$ and describe our problem as a classic Coriolis Effect.

The expression in Eq. (6) describes a spiral-like curve, which is a typical result when a uniform motion in a straight line is seen in a uniformly rotating frame of reference. Eq. (6) can be brought into the classic form of an Archimedean spiral by neglecting the Earth radius $R \ll 1 AU$, which is much smaller than the Sun-Earth distance.

In general, an Archimedean spiral is the curve traced out by a point that moves at constant velocity v along a rod that is rotating about the origin at a constant angular velocity ω. Its equation in polar coordinates is $r = a|\varphi|$, $a = v/\omega > 0$, $-\infty < \varphi < +\infty$, composed of two different branches for positive or negative values of φ.

In our case, for R=0, Eq. (6) reduces to $r' = \sqrt{x'^2 + y'^2} = -\tfrac{1}{2}\alpha$, $\varphi' = -\alpha$, since the α parameter has negative values, thus obtaining the spiral curve

$$r' = \tfrac{1}{2}\varphi', \tag{7}$$

which obeys the general expression since in our case $a = \frac{v}{\omega} = \frac{c}{\omega} = \frac{1/8}{1/4} = \frac{1}{2}$, in our choice of units.

Finally, it is possible to reduce Eq. (6) to a more compact expression, without any approximation. The first part of Eq. (6) can be solved for $\alpha = -2\sqrt{r'^2 - R^2}$, where the minus sign again is chosen because we use a negative α in the parameterization of our curve. Using this last expression and the second part of Eq. (6) we obtain:

$$\sin\varphi' = \frac{\tan\varphi'}{\sqrt{1+\tan^2\varphi'}} = \frac{\frac{1}{2}\alpha\sin\alpha + R\cos\alpha}{\sqrt{\frac{1}{4}\alpha^2 + R^2}} = -\sqrt{1-\frac{R^2}{r'^2}}\sin\alpha + \frac{R}{r'}\cos\alpha. \qquad (8)$$

Since $R/r' \leq 1$, we can set $\sin\beta = \frac{R}{r'}, \cos\beta = \sqrt{1-\frac{R^2}{r'^2}}$ and use trigonometric relations to rewrite Eq.(8) as $\sin\varphi' = \sin\beta\cos\alpha - \cos\beta\sin\alpha = \sin(\beta-\alpha)$, from which we finally deduce:

$$\varphi' = \beta - \alpha = \arcsin\left(\frac{R}{r'}\right) + 2\sqrt{r'^2 - R^2} \qquad (9)$$

which is the most compact expression of the light trajectory in the rotating system, directly connecting the polar coordinates r' and φ' and will also reduce to the spiral of Archimedes for $R \to 0$.

**A sunset-sunrise demonstration**

The discussion presented above suggests a very simple experiment which can be used as an effective in-class demonstration of these rotational effects or even become part of a more structured laboratory activity.

Our experiment is an adaptation of the standard "Coriolis effect – Ball on rotating platform" demonstration,[4] combining together two very basic pieces of equipment from introductory mechanics labs: a rotating platform and an inclined plane which serves as a projectile launcher.

A small metal ball is launched from the inclined plane over the rotating turntable and will represent our beam of light traveling at almost constant velocity in the fixed frame of reference.[5] The Earth is represented on the turntable by a green circle (see Fig.3) and the projectile is launched along a tangential direction to the circle in the fixed frame of reference. This fixed direction is represented by a meter stick attached to the projectile launcher (visible in the upper right corner of the figure). A small orange circle, representing the Sun, is attached to the fixed inclined plane (its position is better identified by the red dots in the figure).

A video camera is mounted on top of the rotating platform and records the view of the rotating observer, as seen in the x'-y' plane. We filmed the motion of the projectile in the rotating frame and then used video editing software to produce a "stroboscopic" picture of the motion.

Figure 3 illustrates one of the pictures we obtained, showing a close resemblance to the spiral curve plotted in Fig. 2 following Eq. (6) or (9). The metal ball is launched when the "Sun" is at position A and then reaches the "observer" at T when the "Sun" is approximately at point B, therefore aligned with the horizon direction as expected.

Although similar rotating platform demonstrations are described by many papers and articles,[6] we are not aware they have ever been used to illustrate our simple sunrise-sunset problem, which could therefore represent a new way to introduce the Coriolis Effect and related topics.

**Relativistic analysis of the problem**

A final point needs to be considered in the analysis of the original problem. We have used concepts of non-relativistic kinematics of rotating systems to study the motion of a beam of light. Should we consider corrections arising from special relativity?

The relativistic treatment of rotating systems and their connection to inertial ones is a topic which is seldom discussed in standard relativity textbooks. We found a complete analysis in a classic book by H. Arzeliès.[7]

In a relativistic approach, when a rigid object is rotating at constant angular velocity ω, we need to ensure that all the linear velocities of all the points of the object do not exceed the speed of light, $v = r\omega < c$, for all the coordinates r of the body, and this is obviously the case of our rotating planet.

Then, at any given instant, an infinitesimal element P of the body can be approximately regarded as an inertial system moving with velocity $v = r\omega$ with respect to the fixed system and standard Lorentz transformations, length contraction and time dilation effects will apply.

In particular, the radial coordinate, perpendicular to the instantaneous velocity of the element P, will not be contracted, i.e., $r' = r$, but any "tangential" length will be affected. For example, the circumference length l is Lorentz contracted:

$$l' = l / \sqrt{1 - \tfrac{v^2}{c^2}} = 2\pi r / \sqrt{1 - \tfrac{(r\omega)^2}{c^2}} = 2\pi r' / \sqrt{1 - \tfrac{(r'\omega)^2}{c^2}} > 2\pi r', \quad (10)$$

where, as before, primed quantities refer to the rotating system as opposed to non-primed ones referring to the fixed system.

Simple Euclidean geometry would therefore not apply in the rotating frame: the circumference to radius ratio would be bigger than the standard 2π factor, as seen from Eq. (10).

On the contrary, the derivation of the light trajectory in rotating systems is not affected at all by relativistic corrections; this is mainly due to the invariance of the radial coordinates $r' = r$ mentioned above.

An alternative derivation of the light trajectory starts by noting that the equation of motion of our beam of light is simply $R = r \sin \varphi$, using polar coordinates in the fixed system (see Fig. 1b) and the connection between polar angles φ and φ' is

$$\varphi = \varphi' + \alpha = \varphi' + \alpha_0 + \omega(t - t_0) = \varphi' + \omega t = \varphi' + \tfrac{1}{4} t \quad (11)$$

The time coordinate in the last equation can be eliminated by noting that in the inertial system the speed of light is fixed to c=1/8 in our units and therefore the distance traveled by the light in time t is:

$$ct = \tfrac{1}{8} t = -\sqrt{r^2 - R^2} = -\sqrt{r'^2 - R^2} . \quad (12)$$

The negative sign comes from the direction of the beam of light and we can exchange the r, r' coordinates as mentioned above. Using Eqs. (11) and (12), we can obtain the trajectory in terms of r' and φ':

$$R = r \sin \varphi = r' \sin(\varphi' + \tfrac{1}{4} t) = r' \sin(\varphi' - 2\sqrt{r'^2 - R^2}), \quad (13)$$

which, solved for φ', gives exactly the same result of Eq. (9) and therefore represents the same solution in our original Eq. (6).

This alternative derivation is totally equivalent to the previous one (and actually simpler), but we emphasize that it is also an exact solution when relativistic effects are taken into account, i.e., it would be correct even if the Earth were spinning very fast, with points on its surface reaching relativistic speeds.

We will leave all further details of the relativistic analysis to the cited textbook,[7] but we can conclude that even in a fully relativistic treatment of the problem the Sun at sunset or sunrise is precisely where we see it: at the horizon and not below or above it!

**Conclusion**

An apparently simple question about phenomena that we witness every day, such as a beautiful sunrise or sunset, can be useful to introduce in-class discussion on rotational frame of references, apparent motion, Coriolis Effects and even more advanced topics in relativity. A very simple demonstration can be easily assembled for further illustrating the problem, which might also lead to a more structured lab activity.

**Acknowledgments**

This research was supported by an award from Research Corporation. The author would like to acknowledge his friend Prof. G. Tonzig, who reported the original statement debated in this paper, and his colleague Dr. J. Phillips for useful discussions.

horizon. The atmosphere actually bends light down toward the surface when a celestial object is near the horizon allowing us to see the Sun or other stars before they would normally be visible if there were no atmosphere.

3. In this way we are over-estimating a little the speed of light in vacuum $c = 1\,AU/8\min = 3.12 \times 10^8\,m/s,$ as opposed to the standard value of $c = 2.99792458 \times 10^8\,m/s,$ but this does not alter the physical explanation of the problem.

4. Physics Instructional Resource Association (PIRA) demonstration 1E30.28, "Coriolis effect - Coriolis ball on turntable" and references therein. See PIRA web page at http://www.wfu.edu/physics/pira/. Video recordings of this demonstration and references are also available at the University of Maryland Physics Lecture Demonstration Home Page: http://www.physics.umd.edu/lecdem/services/demos/demosd5/d5-11.htm.

5. Strictly speaking the rolling motion of a sphere on a horizontal turntable cannot be described just in terms of fictitious forces, due to the rotating frame of reference. Friction is also playing a role, but will not change the trajectory significantly. See the discussion in: Robert H. Romer, "Motion of a sphere on a tilted turntable," Am. J. Phys. 49, 985 (Oct. 1981).

6. See for example: Robert H. Johns, "Physics on a Rotating Reference Frame," Phys. Teach. 36, 178 (Mar. 1998); Richard Andrew Secco, "Coriolis-Effect Demonstration on an Overhead Projector," Phys. Teach. 37, 244 (Apr. 1999); Andreas Wagner, et al., "Multimedia in physics education: a video for the quantitative analysis of the centrifugal force and the Coriolis force," Eur. J. Phys. 27, L27 (2006).

7. H. Arzeliès, Relativistic Kinematics, Pergamon Press, Oxford, 1966. See chapter IX – The Rotating Disc- for a complete relativistic analysis of light paths in rotating frames.

**Gabriele U. Varieschi** is an Associate Professor of Physics at Loyola Marymount University. He earned his Ph.D. in theoretical particle physics from the University of California at Los Angeles. His research interests are in the area of astro-particle physics and cosmology.

**Department of Physics, Loyola Marymount University, 1 LMU Drive, Los Angeles, CA 90045; gvarieschi@lmu.edu**

**FIGURES:**

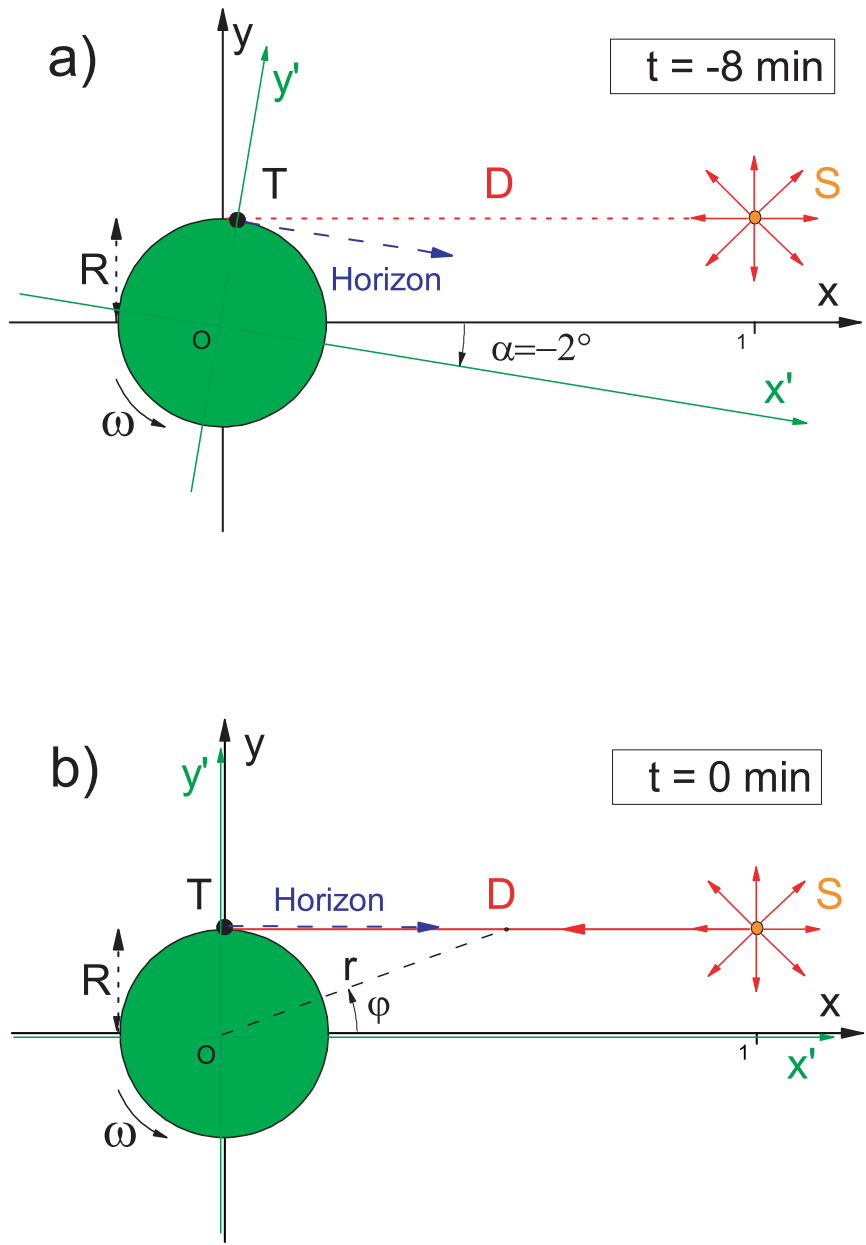

Fig. 1. The view from the fixed inertial system: a) The Sun is emitting its "last" light eight minutes before sunset (t = -8 min). At this time the Sun is well above the horizon direction, by an angle of two degrees (greatly exaggerated in the figure). b) The situation at sunset (t = 0 min). The horizon direction for the terrestrial observer at T is now aligned with the direction of the incoming "last" light from the Sun.

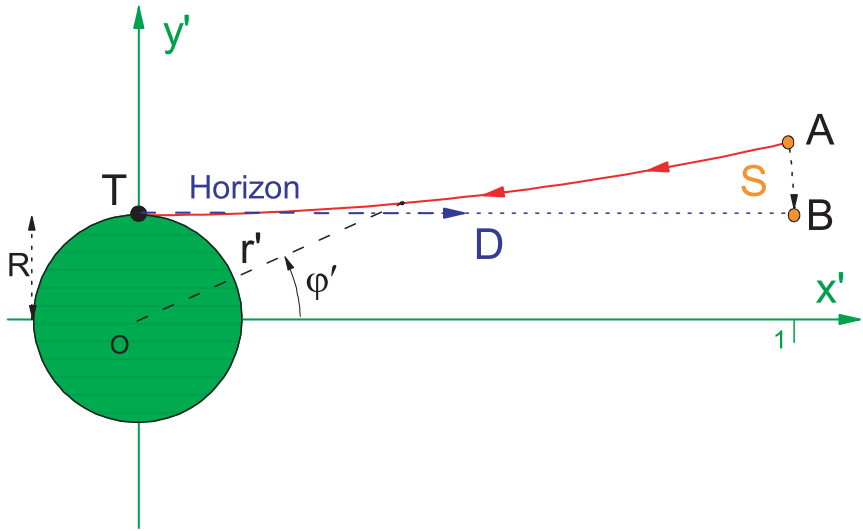

Fig. 2. The view from the Earth's rotating system. The "last" light is emitted from the Sun at position A (t = -8 min) and appears to be following a spiral-like path towards the terrestrial observer at point T. However, the light is perceived as coming from the horizon direction at sunset (t = 0 min). At the same time the Sun is at position B, as seen from the rotating system, therefore perfectly aligned with the horizon. The distances D and R in the figure are not up to scale.

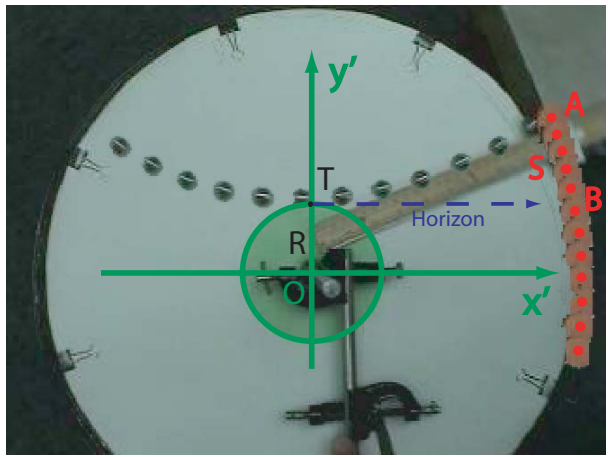

Fig. 3. A stroboscopic picture of the motion produced by our experimental apparatus, as recorded by a video camera rotating together with the turntable. The resulting curved trajectory of the metal sphere, from point A to point T, resembles closely the light path described in Fig. 2. The position of the Sun, in each frame of the video recording, is indicated by the red dots in the figure.